*Semi-relativistic hydrodynamics of three-dimensional and low-dimensional quantum plasma*
Pavel Andreev, Alexey Ivanov, Leonid Kuz'menkov,
Lomonosov Moscow State University, Russia

**Abstract**
Contributions of the current-current and Darwin interactions *and* weak-relativistic addition to kinetic energy in the quantum hydrodynamic equations are considered. Features of hydrodynamic equations for two-dimensional layer of plasma (two-dimensional electron gas for instance) are described. It is shown that the force fields caused by the Darwin interaction and weak-relativistic addition to kinetic energy are partially reduced. Dispersion of three- and two-dimensional semi-relativistic Langmuir waves is calculated.

Theory of special relativity shows that we can consider different regimes of velocity of particles. If velocity of a particle v is much smaller of the speed of light $c$ we have the non-relativistic case. When velocity of a particle v is getting closer to the speed of light same relativistic effects have to be included. This is the semi-relativistic (or weak relativistic) regime. Which is intermediate regime between the non-relativistic and the relativistic limits. Since the Dirac equation has been developed for a particle in an external electromagnetic field, we can't rigorously derive equations for collective many-particle evolution. Consequently we use semi-relativistic approach, when many-particle Schrodinger equation can be used. Considering spinless semi-relativistic Breit Hamiltonian we have following interactions: kinetic energy of each particles, potential energy of each charge in external field, the Coulomb interaction between charges of the system, which are non-relativistic terms, and next semi-relativistic addition to kinetic energy, the current-current interaction (Biot-Savart law) for interaction between moving charges, the Darwin interaction (which is a quantum-relativistic effect) of charges with external electric field, and the Darwin interaction between charges, when a charge is under influence of electric field created by other charges [1], [2].

Dispersion of two-dimensional Langmuir wave is considered along with the three-dimensional Langmuir waves. Thus we have to explain meaning of notion "two-dimensional", since this notion has been used in two different meanings in physical literature. Usually we have deal with three-dimensional samples. We consider propagation of plane waves (one-dimensional perturbations), cylindrical waves (two-dimensional perturbations), and spherical waves (three-dimensional perturbations). However, these days technology allows to manipulate with low dimensional samples, plane-like objects – two dimensional samples, up to one atom layer structures as graphene. Therefore we should be conscious and careful using notions two-dimensional or one-dimensional medium. In this paper we consider two-dimensional plane of plasma.

The first and crucial step in derivation of equations for collective evolution of quantum systems is definition of many-particle concentration as the quantum mechanical average of corresponding operator [3]-[6].

When we have derived equations of many-particle quantum hydrodynamics directly from many-particle Schrodinger equation, we have them in integral form. For instance let us present the Euler equation for non-relativistic plasma in the self-consistent field approximation for illustration of this statement

$$mn(\partial_t + \mathbf{v}\nabla)\mathbf{v} + \nabla p - \frac{\hbar^2}{2m}\nabla\left(\frac{\Delta n}{n} - \frac{(\nabla n)^2}{2n^2}\right)$$
$$= en\left(\mathbf{E}_{ext} + e\nabla\int\frac{n(\mathbf{r}',t)}{|\mathbf{r}-\mathbf{r}'|}d\mathbf{r}'\right) + \frac{e}{c}n[\mathbf{v},\mathbf{B}_{ext}], \quad (1)$$

where $n = n(\mathbf{r},t)$, $\mathbf{v} = \mathbf{v}(\mathbf{r},t)$, $\mathbf{r} = \{x,y\}$, $p$ is the isotropic pressure, the quantum Bohm potential is presented by terms proportional the reduced Plank constant $\hbar$, both these terms are presented in the left-hand side of equation (1). They have kinematic nature and are not related to interaction explicitly. However, structure of the wave function depends on interaction and quantum Bohm potential in formula (1) was obtained form approximately non-interacting particles. The right-hand side of equation (1) contains interparticle interaction and interaction of particles with external electromagnetic fields. In many cases plasmas are considered in the presence of constant uniform magnetic field. In tokomaks we have more complicate structure of external magnetic field. The first and third terms describe interaction with the external electromagnetic field. The second term gives contribution of the Coulomb interaction and has integral form.

However in semi-relativistic approximation the Euler equation has rather more complicate form, since basic Hamiltonian contains more interactions. Reduction of the Euler equation for three-dimensional samples was presented in Refs. [7] and [8], while Ref. [8] does not contain contribution of the Darwin interaction.

Semi-relativistic effects give contribution as in kinematic effects and in interparticle interaction. The quantum Bohm potential receives considerable contribution. Semi-relativistic part has complex tensor structure containing combination of two vectors of velocity field and non-relativistic part of quantum Bohm potential, it also contain combination of particles concentration and spatial derivatives of velocity field, and group of terms, which does not contain contribution of velocity field including only concentration and spatial derivatives of concentration up to the fourth (see Ref. [7] formula (20) for explicit form). Factor $1/c^2$ is the mark showing that terms have semi-relativistic nature, so we can recognize them even when do not contain contribution of velocity field.

Commutation of the operator of semi-relativistic part of kinetic energy with operator of the Coulomb interaction leads to two kind of terms in the force field. One of them is quantum generalization of classic semi-relativistic force (third line of equation (14) in Ref. [7]). The second one has quantum relativistic nature and has following form

$$\mathbf{F}_{s.r.} = -\frac{\hbar^2}{4m^2c^2}\nabla(\nabla\mathbf{E}) . \quad (2)$$

It has structure similar to the force field caused by Darwin interaction given by

$$\mathbf{F}_D = \frac{\hbar^2}{4m^2c^2}\nabla(n\nabla\mathbf{E}) = \frac{\hbar^2}{4m^2c^2}\left(n\nabla(\nabla\mathbf{E}) + \nabla n\nabla\mathbf{E}\right). \quad (3)$$

They partly cancel each other

$$\mathbf{F}_D + \mathbf{F}_{s.r.} = \frac{\hbar^2}{4m^2c^2}\nabla n\nabla\mathbf{E}, \quad (4)$$

so they give no contribution in linear approximation if we have homogeneous medium in equilibrium state.

As it expected the current-current interaction gives Lorentz force density. However it is not only one contribution of the current-current interaction. Commutation of the operator of current-current interaction with operators of the non-relativistic kinetic energy and the potential energy of Coulomb interaction gives four extra terms. One of this terms appears to have quantum nature, it explicitly contains square of the Plank constant. As trace of current-current interaction is the presence of corresponding Green function, which is a tensor having form

$$G^{\alpha\beta}(\mathbf{r},\mathbf{r}') = \frac{\delta^{\alpha\beta}}{|\mathbf{r}-\mathbf{r}'|} + \frac{r^\alpha r^\beta}{|\mathbf{r}-\mathbf{r}'|^3} . \quad (5)$$

Quantum kinetic consideration of the three-dimensional Langmuir waves in semi-relativistic approximation including the Darwin interaction, but with no account of the semi-relativistic addition to kinetic energy was presented in Ref. [9].

Considering dynamics of small longitudinal high frequency perturbations of equilibrium homogeneous plasma in absence of external fields assuming that ions are motionless we find spectrum of semi-relativistic quantum Langmuir waves for three dimensional sample

$$\omega^2 = \omega_{Le,3D}^2\left(1 - \frac{5T_0}{2mc^2}\right) + 3v_s^2 k^2 + \frac{\hbar^2 k^4}{4m^2}\left(1 - \frac{\hbar^2 k^2}{2m^2 c^2}\right), \quad (6)$$

where we included that $\mathbf{F}_D + \mathbf{F}_{s.r.}$ gives zero contribution. The first group of terms in formula (6) consists of two parts: the Langmuir frequency and semi-relativistic thermal shift. The second term is the contribution of thermal pressure. Here we applied the fact that equation of state for relativistic and non-relativistic ideal classical gas is the same $p = nT$. The last group of terms has quantum nature. It contains two terms: quantum Bohm potential and semi-relativistic part of the quantum Bohm potential.

Considering same problem for two-dimensional plasma layer we obtain

$$\omega^2 = \omega_{Le}^2\left(1 - \frac{2T_0}{mc^2}\right) + 3v_s^2 k^2 + \frac{\hbar^2 k^4}{4m^2}\left(1 - \frac{\hbar^2 k^2}{2m^2 c^2}\right)$$
$$- \frac{\omega_{Le}^2}{2k^2 c^2}\left(\omega_{Le}^2 + 3v_s^2 k^2 + \frac{\hbar^2 k^4}{4m^2}\right), \quad (7)$$

where we have two-dimensional Langmuir frequency

$$\omega_{Le}^2 = \frac{2\pi e^2 n_0 k}{m},$$

here $[n_0] = cm^{-2}$ equilibrium two-dimensional concentration of particles. This result is also included that $\mathbf{F}_D + \mathbf{F}_{s.r.}$ gives zero contribution. The first line in formula (7) has meaning similar to terms in formula (6). Including the fact that particles in two dimensional sample have one degree of freedom less we have another coefficient in front of the second term of the first group of terms. The last group of terms in formula (7) is caused by the current-current interaction. In three-dimensional case, this group of terms does not appear due to properties of the Green function.